# Replica-Averaging: An algorithm to study mechano-reactive processes for polymers under flow conditions

Sagar Kania[1], Anh H. Nguyen[1], Alparslan Oztekin[1], and Edmund Webb III[1,*]

[1] Department of Mechanical Engineering and Mechanics, Lehigh University, Bethlehem, Pennsylvania 18015

[*] Corresponding author e-mail: ebw210@lehigh.edu

## Abstract

A new method based on quasi-independent parallel simulations approach, replica-averaging, has been developed to study the influence of flow on mechanical force-mediated polymer processes such as denaturation and breaking of bonds. This method considerably mitigates the unphysical prediction of force-mediated events inherent in Brownian dynamics (BD) polymer chain simulations that employ instantaneous force profile-based criteria to identify the occurrence of such events. This inaccuracy in predicting force-mediated event kinetics is due to high fluctuations of the instantaneous force profile around the average force. Replica-averaging reduces such high fluctuation effects by computing a force profile that faithfully represents the average force profile of the polymer chain conformation, which is then used to predict reactive events. For transient conformation conditions, the replica-averaged method more accurately predicts mechano-reactive kinetics than the time-averaged method, typically employed to reduce the unphysical prediction of force-mediated events in BD simulations. Further, the influence of the proposed replica-averaging method parameters on the accuracy of predicting the true average force profile along the polymer is discussed.

# I. Introduction

Mechanical force lowers the activation energy barrier for many transitions from one stable state to another. The extent to which it is reduced may also depend on the loading rate[1–4]. Processes controlled by such transitions are referred to here as mechano-reactive. Polymers exhibit mechano-reactive processes that underpin molecular functionality, such as denaturation and breaking of bonds; thus, there is significant interest in understanding mechano-reactive polymer processes[5–8]. Such knowledge enhances the understanding of biopolymeric system functionality and degradation of polymers used for drag reduction or enhanced oil recovery fluids[9–13]. A prominent mechanism for inducing mechanical stress in polymers is the hydrodynamic force of fluid flows. Generally, the time and length-scale of fluid flows regulating mechano-reactive polymer processes are inaccessible for high-resolution simulation techniques like molecular dynamics. Therefore, Brownian dynamics (BD) simulations of coarse-grained polymer chain models (bead-rod or bead-spring) are employed to study the influence of flow-induced changes in the polymer chain conformation on the kinetics of mechano-reactive polymer functionality.

Mechano-reactive event kinetics for polymers under varied flow conditions can be estimated using the instantaneous spring or rod force, which is one term in the overdamped Langevin equation that governs particle dynamics in BD simulations[12,14–16]. Such an approach assumes that the instantaneous spring or rod force represents the average chemical scale force (i.e., force on a bond or force on a molecular cluster) with reasonable accuracy. In some studies, instantaneous spring force values are directly compared to the bond rupture force to detect bond-breaking events[12,15,16]. In other studies, the instantaneous forces are employed to estimate an event's transition rate via the Bell model, which has been widely used to characterize changes in the free energy profile caused by mechanical stress[4,14]. However, predicting the kinetics of mechano-reactive polymer processes using instantaneous force from BD simulations can generate incorrect results due to the stochastic term in the governing Langevin equation, coupled with the coarse-graining of time[12,17]. To be clear, average values in a system, including spring force as a representative of chemical scale force, are well captured by the simulation method[17,18]. However, the coarse grain BD method introduces fluctuations in spring force that deviate from the average force much more than would be observed at the atomistic or chemical scale, which is germane to reaction kinetics[12,17]. Similar to prior authors studying mechano-reactive events in BD simulations, in the present study, such high fluctuation is characterized as "unphysical"[17,19].

One way to circumvent unphysical fluctuations in the spring or rod force inherent in coarse grain stochastic BD simulations is to use the normalized tensile profile to predict tensile-force induced polymer events[17]. This method is based on the concept that the probability distribution of instantaneous force follows a Gaussian distribution with the average force as the mean of the distribution. Sim et al. employed the normalized tensile profile to study the scission dynamics of polymer chains in elongational flow[17]. To obtain a normalized tensile force profile for various polymer chain conformations in elongational flows, those authors used an analytical expression for average force and standard deviation of the tensile force. By employing this normalized tensile force, they showed that the location of bond rupture along the polymer chain depends on the elongational flow strain rate[17]. Moreover, it is possible to filter the unphysical fluctuation effects for free-draining bead-rod polymer chains by employing an instantaneous, deterministic tension derived by Schieber and Obasanjo[20]. The authors divided the constraint tension into an instantaneous, deterministic segmental tension and a stochastic part. Though both methods successfully negate the unphysical fluctuation effect, they have limitations. The normalized tensile force method requires analytical expressions related to tensile force, which are difficult to estimate when polymeric chains are under complex flow conditions[17]. Furthermore, the instantaneous, deterministic tension expression is not valid for the bead-spring polymer model; more importantly, it is not applicable when hydrodynamic interaction significantly influences the polymer dynamic behavior[17,20], such as for polymers under poor solvent conditions[21].

Another way of overcoming the BD simulation's unphysical fluctuation effect is by using a numerically computed polymer chain confirmation's quasi-equilibrium (average) force profile that minimizes algorithm-induced fluctuations of the rod or spring force. In the absence of algorithm effects, stochastic fluctuations around the quasi-equilibrium rod/spring force for a transient polymer configuration follow Gaussian distributions[17]; we identify them as quasi-equilibrium Gaussian distributions. Furthermore, it has been illustrated that the actual variance - i.e., the variance free from the artifacts of BD simulations - of the quasi-equilibrium Gaussian distributions is too small to significantly influence mechano-reactive polymer event kinetics[17]. This suggests that predicting force-induced polymer events using the quasi-equilibrium force profile is valid.

To compute the quasi-equilibrium force profile for a transient polymer configuration, it is essential to take sufficient uncorrelated samples to ensure that the computed average force profile reduces algorithm-induced fluctuations until their influence on predicted mechano-reactive kinetics is minimized (i.e., until the true average spring force controls kinetics). For polymers under varied flow conditions, the quasi-equilibrium force profile is usually computed as a time average of the instantaneous spring or rod force profile over a sufficient sampling time during BD simulations[19]. Generally, to achieve a well-bounded prediction of the average force, the instantaneous force must be averaged over a time that may prove problematic in scenarios where polymer conformation changes rapidly. For example, the average might represent a convolution of conformations structurally distinct from one another. The force obtained may result in inaccurate predictions of the flow effect on force-induced polymer event kinetics.

Here, we introduce a new simulation method, replica-averaged spring force sampling, to compute average force profiles that faithfully represent a chain conformation, even in highly dynamic scenarios. The replica-averaged method is based on the quasi-independent parallel simulation approach, extensively used in path sampling techniques such as Weighted Ensemble[22]. Further, we characterize conditions in which the replica-averaged method would outperform the time-averaged method for analyzing mechano-reactive polymer functionality in varied flow conditions. We discuss the dependence of the replica-averaged method's accuracy in predicting average spring force on its parameters. We have employed collapsed polymers (polymers under bad solvent conditions) to compare time-averaged and replica-averaged force sampling. Polymers under bad solvent conditions were chosen for our study because they exhibit very rapid conformation transitions in a high shear rate flow[23]. Therefore, it is difficult for sampling methods to accurately track the average spring force representative of the collapsed polymer system's transient dynamics.

## II. Methodology

**Collapsed polymer model and Brownian Dynamics (BD) simulation**

Typically to study the dynamic behavior of collapsed polymers in solutions, they are represented as a series of beads connected by springs, i.e., a bead-spring polymer model[21,24–26]. In such models, the spherical beads act as discretized sources of friction, springs that connect neighboring beads represent the backbone of the polymer chain, and self-association of the polymer is described via interaction between non-adjacent beads. Our bead-spring model is composed of $N = 50$ beads of radius $a$ and interacting through a potential $U = U_s + U_{LJ}$. $U_s$ accounts for the connectivity of the chain and is given as $U_s = \frac{\kappa}{2}\sum_{i=1}^{N-1}(r_{i+1,i} - 2a)^2$, where $r_{i+1,i}$ is the distance between adjacent beads, and $a$ is the bead radius. The spring constant is taken to be $\kappa = 400 k_b T/a^2$ ($k_b$ = Boltzmann constant and $T$ = Temperature), which limits the bead-spring polymer model to stretch beyond its contour length to a negligible level[26]. The Lennard-Jones potential represents the self-association of the polymer $U_{LJ} = u \sum_{ij}[(2a/r_{ij})^{12} - 2(2a/r_{ij})^6]$, where $r_{ij}$ is the distance between the $i^{th}$ and $j^{th}$ bead and $u$ determines the depth of the potential. To model a collapsed polymer[21], $u$ is considered as $1.0 k_b T$.

For a spatially homogenous velocity gradient and in the presence of hydrodynamic interaction (HI), the stochastic equation governing the evolution of the position vector, $\mathbf{r}_I$, for the i$^{th}$ bead, is given by[27]:

$$\mathbf{r}_i^{t+\Delta t} = \mathbf{r}_i^t + \left[\mathbf{v}_\infty(\mathbf{r}_i^t) - \frac{1}{k_bT}\sum_{j=1}^N \underline{D}_{ij}(\mathbf{r}_i^t, \mathbf{r}_j^t) \cdot \nabla_{\mathbf{r}_j^t} U(t)\right]\Delta t + \mathbf{R}_i(\Delta t) \quad (1)$$

where, $\mathbf{r}_i^t$ and $\mathbf{r}_i^{t+\Delta t}$ are the position of i$^{th}$ bead at time step t and t+$\Delta$t, respectively. For BD simulations, we use a time step $\Delta$t of $10^{-5}\tau$, where $\tau$ is the single bead diffusion time $\tau = 6\pi\eta a^3/k_bT$ ($\eta$ is solvent viscosity). $\mathbf{v}_\infty(\mathbf{r})$ is the undisturbed solvent velocity and $\mathbf{R}_i(\Delta t)$ is a random displacement whose average is 0 and variance-covariance is $\langle \mathbf{R}_i(\Delta t)\mathbf{R}_j(\Delta t)\rangle = 2\underline{D}_{ij}\Delta t$. Hydrodynamic interaction among beads is manifested in $\underline{D}_{ij}$ (diffusion tensor), which is given by the Rotne-Prager-Yamakawa approximation[27–29]:

$$\underline{D}_{ij} = \frac{k_bT}{6\pi\eta a}\underline{\mathbb{I}}; \quad i = j \quad (2)$$

$$\underline{D}_{ij} = \frac{k_bT}{8\pi\eta r_{ij}}\begin{cases} \left(1 + \frac{2a^2}{3r_{ij}^2}\right)\underline{\mathbb{I}} + \left(1 - \frac{2a^2}{r_{ij}^2}\right)\frac{\mathbf{r}_{ij}\mathbf{r}_{ij}}{r_{ij}^2} & r_{ij} \geq 2a \\ \frac{r_{ij}}{2a}\left[\left(\frac{8}{3} - \frac{3r_{ij}}{4a}\right)\underline{\mathbb{I}} + \frac{r_{ij}}{4a}\frac{\mathbf{r}_{ij}\mathbf{r}_{ij}}{r_{ij}^2}\right] & r_{ij} < 2a \end{cases}; i \neq j \quad (3)$$

where i, j denote beads, $\mathbf{r}_{ij} = \mathbf{r}_j - \mathbf{r}_i$, and $\underline{\mathbb{I}}$ is the identity matrix. Hereafter, we use the bead radius (a), thermal energy ($k_bT$) and single bead diffusion time ($\tau = 6\pi\eta a^3/k_bT$) as characteristic scales for length, energy, and time, respectively.

We performed BD simulations of collapsed polymers under quiescent and shear flow conditions with a shear rate equal to $\dot{\gamma}$. For quiescent condition $\mathbf{v}_\infty(\mathbf{r})$ equals zero, whereas when polymers are subjected to shear flow $\mathbf{v}_\infty(\mathbf{r}_i) = \dot{\gamma} z_i \hat{\mathbf{x}}$, where $z_i$ is the z-component of the position vector of i$^{th}$ bead, $\hat{\mathbf{x}}$ is the unit vector in the flow direction. Therefore, for our simulations, the z is the velocity gradient direction (crossflow direction), and the x is the streamwise direction. To study the flow-induced rapid conformation transitions of collapsed polymer, we chose $\dot{\gamma} = 30$, significantly greater than the threshold shear rate required to unravel the collapsed polymer model studied here[21].

**Computing Instantaneous spring force (ISF), Time-averaged spring force (TSF), and Replica-averaged spring force (RSF)**

During BD simulations, bead positions at each time step are computed from Eq. 1. From the bead positions, ISF of i[th] spring is computed as[12,16,17]:

$$\text{ISF}_i(t) = \kappa(r_{j,j+1}(t) - 2a) \tag{4}$$

where, $r_{j,j+1}$ is the distance between two beads (j and j + 1) connected by the i[th] spring, $\kappa = 400$ is the spring constant, and 2a is the zero-force separation distance. Time average spring force (TSF) uses a sample of force for each spring, taken each time step; after a designated sampling time S, a mean is computed over all samples and used as the representative force value for the given spring over the duration S. For TSF sampling, data can be taken less frequently, but a typical practice employed to compute TSF sampling is to use a frequency of one simulation time step[19].

For replica average spring force (RSF) sampling, at a point in time when the force on each spring will be sampled, the simulation of a given chain is halted, and $N_R$ replicas of the current conformation (i.e., the parent chain conformation) are formed. Each replica is assigned a unique random number seed used for that replica's subsequent BD simulation of duration S. After time S, the force for each spring in the chain is sampled across all $N_R$ replicas to compute a replica-averaged force for each spring. Note that only a single force value (for each spring) is taken from each replica; force is not periodically sampled over duration S – it is only sampled at the end of the duration. For this reason, when referring to RSF sampling, S is described as an evolution time. This is different from TSF sampling, for which S is sampling time. After replica averaged spring forces have been computed, the original parent chain is continued on its BD simulation trajectory until the next force sampling occurs.

## III. Results & Discussion

**Dependence of replica-average and time-average method performance on their parameters**

The probability distribution function (PDF) for force on a spring in a bead spring polymer model quantitatively assesses how different sampling techniques affect predictions of spring force[17]. A well-characterized condition for a collapsed polymer is zero flow when the polymer is in an equilibrium globular conformation, in which springs between adjacent beads oscillate around a well-defined magnitude of near-zero force. For the middle spring in chains here, i.e., the 25th spring, the average non-dimensional force is 0.55, and this value gives accurate mechano-reactive kinetics[17]. For polymer globules at zero flow, Fig. 1 illustrates the PDF of ISF, TSF for $S = 0.02$, and RSF for $N_R = 50$ and $S = 0.02$. Results for quiescent flow conditions are independent of the spring location (see Supplementary Material); therefore, for this analysis, we only show the force PDF for the middle spring (i.e., 25th spring). Note that $10^4$ data points were used to form each PDF in the figure, but doubling the amount of data did not change the results. Figure 1 shows that sampling spring force via ISF data results in contributions that deviate significantly from the well-defined average; for example, the largest magnitude force values from ISF sampling are over 100x the average. Nearly 90% of force samples from ISF have a magnitude over 10x the average. In short, sampling via ISF is subject to BD fluctuation effects that could predict mechano-reactive events even when the average force along the chain is significantly less than the threshold force required to induce such events. Figure 1 further illustrates that the range of a predicted force is suppressed with the TSF and RSF methods of sampling.

All data in Fig. 1 suggest that the force PDFs obtained via ISF, TSF, and RSF sampling for collapsed polymers in quiescent flow exhibit a Gaussian distribution. Our simulation results indicate that the mean of those Gaussian distributions is unaffected by time-averaging or replica-averaging of spring force. Figure 2 shows the standard deviation for fitted distributions as a function of S, i.e., sampling time for TSF or evolution time for RSF; results are also presented for varying $N_R$ for RSF. To illustrate the time scale over which BD-induced force fluctuations decay, the inset of Fig. 2 shows the ISF autocorrelation function, and it shows decorrelation is complete around S = 0.02. This provides context for the main panel of Fig. 2; the dashed vertical line indicates S = 0.02. All curves in the main panel show standard deviation decreases with increasing S. For TSF sampling, the decrease is significantly less abrupt than is observed for any of the RSF data sets. Figure 2 shows for all RSF data that the drop in standard deviation is steep over the first half of the decorrelation time observed in the inset. TSF sampled data show a gradual decrease in the standard deviation over a timescale larger than the decorrelation time observed in the inset. This is because TSF sampling retains a memory of any large fluctuations in force in the form of samples in the data set. Despite this, Fig. 2 illustrates the standard deviation for TSF sampling is halved by the end of the decorrelation time. Considering the standard deviation from RSF sampling at S = 0.02, for $N_R$ = 10, $\sigma$ = 6.7 and, for $N_R$ = 50, $\sigma$ = 3.3. For TSF sampling to achieve a similar reduction in $\sigma$ (6.7 or 3.3), sampling time must be at least S = 0.07 or S = 0.36, respectively.

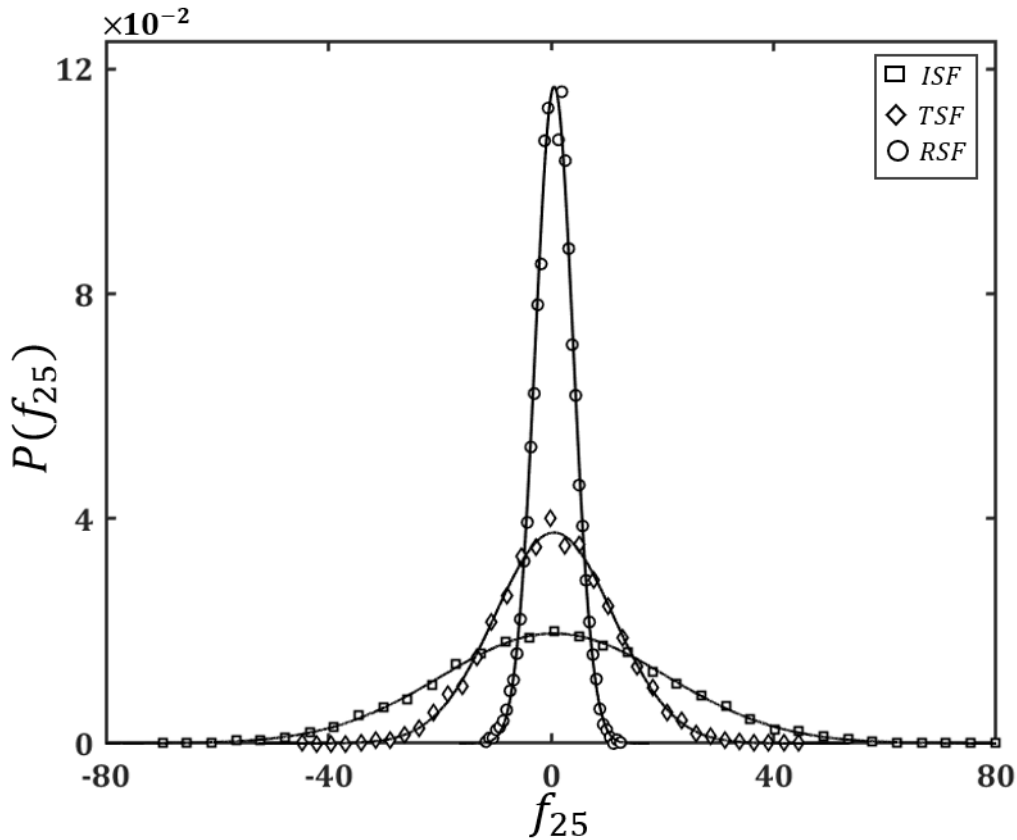

**Figure 1**: The symbols reflect the middle ($25^{th}$) spring force normalized probability distribution ($P(f_{25})$) in the absence of shearing for collapsed polymers; the data are well fit with a Gaussian distribution (lines). Square symbols represent $P(f_{25})$ of instantaneous spring force (ISF). Diamond and circle symbols illustrate time-averaged spring force (TSF) and replica-averaged spring force (RSF) $P(f_{25})$ at $S = 0.02$, respectively; for RSF, $N_R = 50$.

The significant decrease in standard deviation exhibited over the first half of the ISF decorrelation time for RSF sampling depicts that the memory effect present in TSF sampling is removed for RSF. Suppose replicas are created at a moment when the force exhibits a significant BD-induced fluctuation. In that case, each replica launches on a distinct but statistically equivalent trajectory, quickly decorrelating from the starting state. Thus, the force value obtained at evolution time S is more tightly bound around the true average. Data for RSF sampling in the main panel of Fig. 2 also show minimal decreases in standard deviation beyond $S = 0.02$. Standard deviation at $S = 0.02$ decreases by 27% when the number of replicas is doubled from 10 to 20. Doubling again to $N_R = 40$ decreases standard deviation by 24%, in accord with $\sigma \sim 1/N_R^{1/2}$.

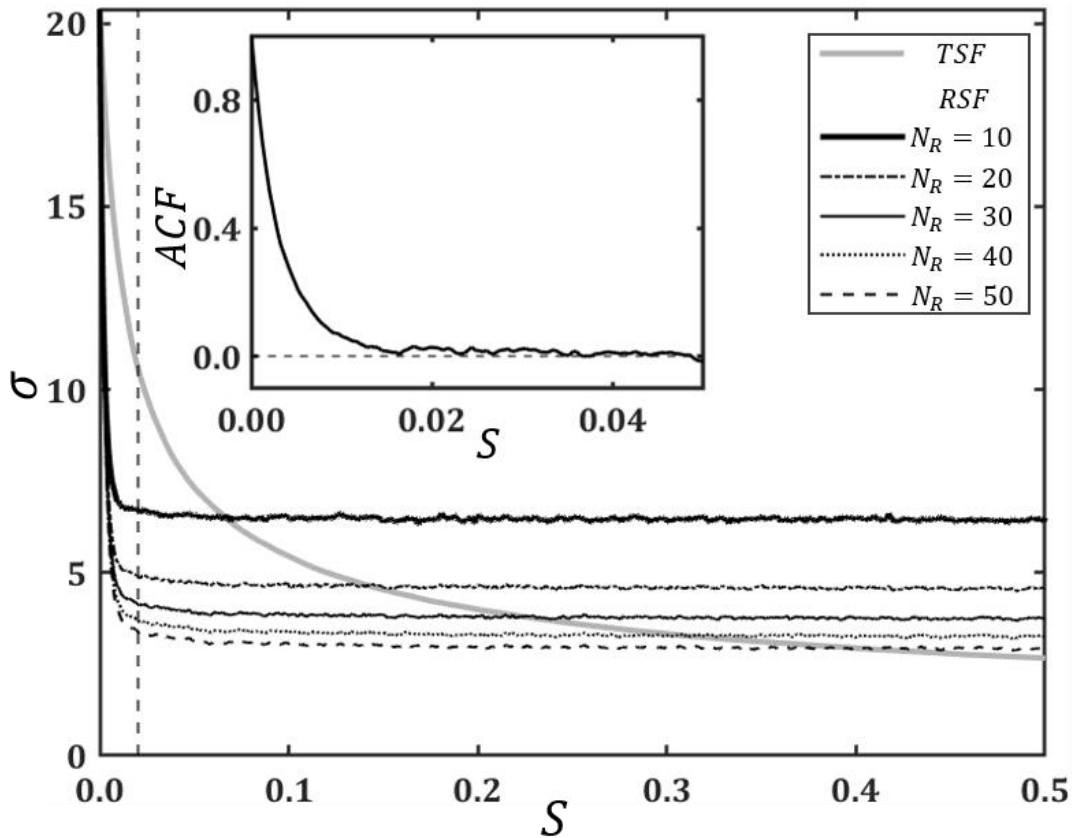

**Figure 2**: The standard deviation (σ) vs. S for RSF and TSF sampling for the $25^{th}$ spring (middle spring). For RSF, data for different numbers of replicas ($N_R$) are shown. The inset shows the instantaneous spring force autocorrelation function (ACF).

For the quiescent condition, time averaging is sufficient - indeed superior, because the standard deviation for predicted force can be reduced to a level comparable to that obtained with RSF sampling at a reduced computational cost. However, to make TSF predictions of the average force comparably accurate to RSF, it is necessary to obtain time averages in the simulation for order S = 0.1 - 0.4. This presents no challenge for an equilibrium, zero flow condition, in which a polymer globule will rotate and diffuse on some characteristic time scale. However, the average separation distance between adjacent beads and the spring force is constant in time. The benefit of RSF sampling is more evident for highly dynamic situations in which polymer conformation is changing rapidly with time.

**Performance of RSF sampling in flow conditions**

When sampling spring forces in BD simulations to be used in mechano-reactive kinetic models, the goal is to obtain a value that is tightly bounded around the average spring force because the average force is what accurately determines kinetics. Further, the value obtained should adequately reflect the polymer conformation at sampling times. Figure 2 shows that the TSF and RSF sampling methods accurately predict the quasi-equilibrium (average) spring force profile experienced by the current polymer conformation, given that sufficient sampling is performed. This requires that any form of averaging is done over conformations that preserve the parent chain conformation; sampled conformations can only be negligibly different from the parent chain conformation. In transient conformation conditions of potential interest to mechano-reactive modeling[30–33], TSF sampling may require sampling time long enough that the preceding requirement cannot be met, leading to inaccurate estimates of the quasi-equilibrium force profile along with the parent chain conformation. Thus, RSF sampling over short times may be motivated.

A condition in which polymers exhibit rapid conformational transitions is for collapsed polymers in shear flows above a transitional, or threshold, shear rate[21,34]. At a relatively low shear rate, collapsed polymers persist in the globular conformation[34]. However, in shear rates greater than the threshold value, polymers exhibit rapid transitions from a globular conformation to a partially or fully elongated conformation and then back to a collapsed globular form[34]. These transitions repeatedly happen in such flows, and their frequency increases as the shear rate increases beyond the threshold value[34,35]. Because of such rapid conformation transitions, high shear rate flow is a condition for which conserving parent chain conformation may be challenging for any sampling method.

To characterize the dynamic nature of a collapsed polymer in high shear rate flow conditions, we consider here a shear rate of $\dot{\gamma} = 30$, which is well above the lowest rate for dynamic transitions between the globule and elongated state [21]. In such a flow condition, the polymer extension in the flow direction varies extensively, from minimum values comparable to those observed in zero flow conditions to maximum values closer to the contour length[12,35]. This leads to a significant fluctuation in force values observed on any given spring in the chain, but fluctuations are most prominent for the middle spring (here, spring 25). The observation that fluctuations are largest for the middle spring in a chain has been made previously[36]. Figure 3 compares predictions for force on the 25th spring obtained via ISF (S = 0) and RSF sampling for different values of S; an extensive range in predicted values is evident for all curves shown. While some benefit can be seen from RSF sampling in constraining the range of force predicted, this effect is relatively small. That is because Fig. 3 amalgamates many different polymer conformations, such that the range in predicted force in all cases is much more extensive than the reduction in that range produced by RSF sampling. This is further supported by the inset to Fig. 3; assuming distributions in the main panel are skew normal, the inset shows the RSF sampling standard deviation as a function of S for various spring position and $N_R = 10$ and 50. Though σ decreases with S, similar to what was observed for the zero-flow case, the magnitude of σ in all cases remains too large to reflect a single conformation's quasi-equilibrium spring force distribution.

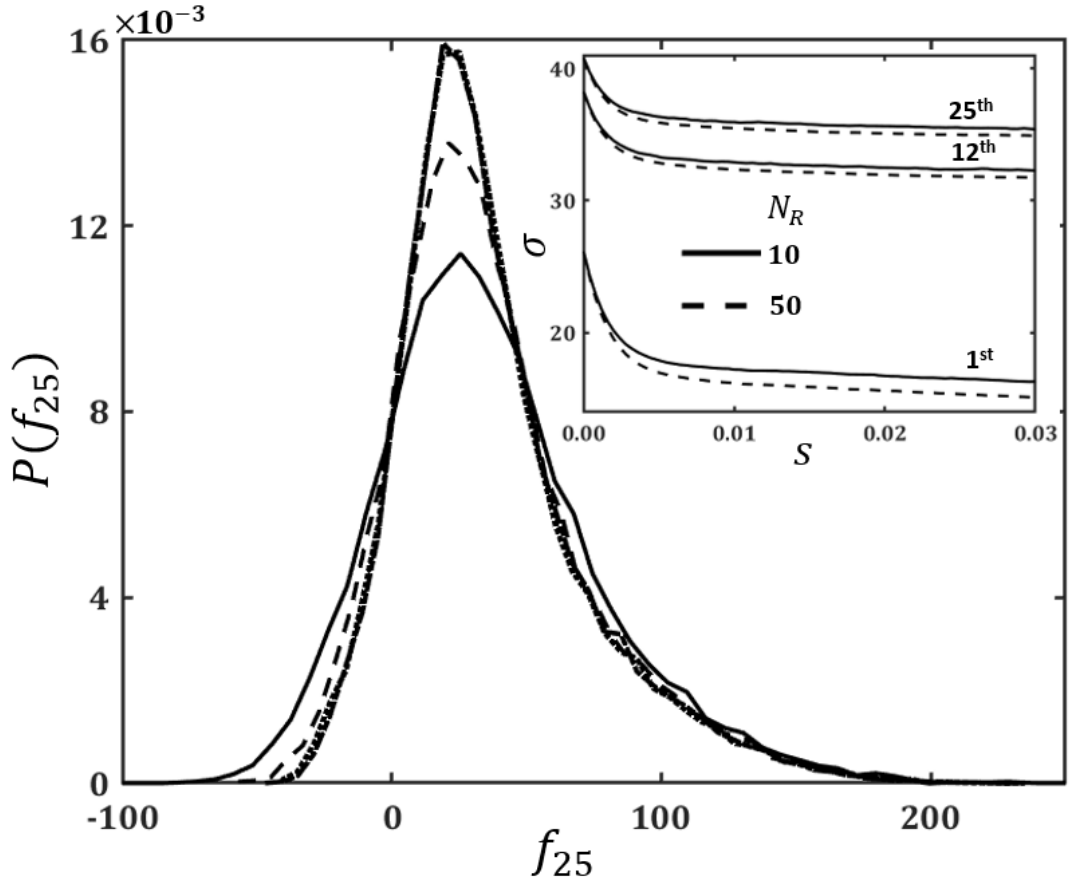

**Figure 3**: 25$^{th}$ spring replica-averaged force normalized probability distribution (P(f$_{25}$)) for different S at $\dot{\gamma}$ = 30. Curves are shown for S = 0 (solid, this is equivalent to ISF sampling), S = 0.0025 (dashed), S = 0.0250 (dotted), and S = 0.0500 (dash-dot). The inset shows the standard deviation (σ) of replica-averaged spring force as a function of S for $N_R$ = 10, 50, and varied spring locations at $\dot{\gamma}$ = 30.

A better way to assess the benefit of RSF sampling is to consider conformation-specific spring force during high shear rate flow. Given the direct relationship between polymer conformation and spring force, this is equivalent to selecting polymer conformations that, after RSF sampling, contribute to a narrow range in the force distributions shown in Fig. 3. For example, for $N_R$ = 50, and using the middle spring force PDF at S = 0.015, all parent polymer conformations can be identified that, after RSF sampling, contributed to the PDF in the range -7.5 < $f_{25}$ < -6.5. Assuming a quasi-equilibrium distribution in $f_{25}$ is present after decorrelation from the parent chain's force profile; Fig. 2 can be used to estimate that the standard deviation for S = 0.015 and $N_R$ = 50 is σ = 3.3. Thus, it is expected that approximately 95% of selected conformations have a true quasi-equilibrium average spring force in the range -14.1 < $f_{25}$ < 0.1. For those selected conformations, Fig. 4(a) plots $f_{25}$ as predicted by ISF sampling (i.e., S = 0); less than 30% of samples are within the identified 95% statistical interval. This analysis was repeated for different force windows in the RSF sampled PDF, i.e., 35.5 < $f_{25}$ < 36.5 and 79.5 < $f_{25}$ < 80.5, and those results are shown in Figs. 4(b) and 4(c), respectively. Results obtained are insensitive to the analysis force window in that ISF sampling consistently produces predictions well outside the expected range of true quasi-equilibrium average spring force. While data presented in Fig. 4 are for the 25$^{th}$ spring, similar observations can be made regardless of spring position (see Supplementary Material).

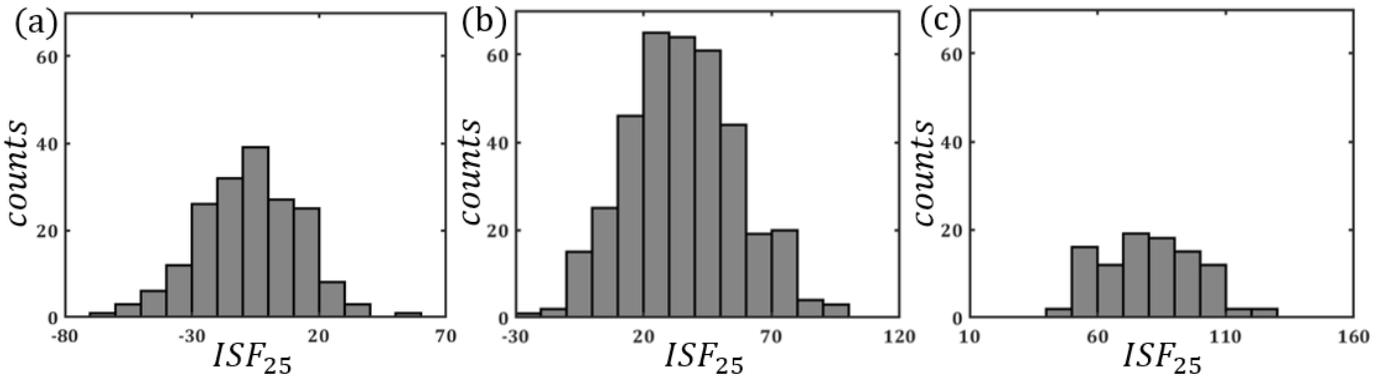

**Figure 4**: 25$^{th}$ spring instantaneous force (ISF$_{25}$) distribution for conformations whose middle spring's (25$^{th}$ spring) RSF for S = 0.015 and $N_R$ = 50 is in the range (a) -7.5 < $f_{25}$ < -6.5, (b) 35.5 < $f_{25}$ < 36.5 and (c) 79.5 < $f_{25}$ < 80.5.

For each force window analyzed in Fig. 4, Fig. 5 shows the computed standard deviation for the ISF sampled data; note again that this is the same as S = 0 for RSF sampling and the value obtained, regardless of force window, is σ ~ 20. For each analyzed force window, selected conformations were subject to RSF sampling using varying $N_R$ and for increasing S. All plots show that increasing S results in an initial steep decline in σ, and, for S > 0.005, σ decreases more gradually with increasing S. Observed behavior in σ versus S is insensitive to the force window analyzed. Similar to what was seen for the quiescent state in Fig. 2, evolution time on the order of S = 0.010 to 0.015 provides a good balance between constraining added computational cost while significantly narrowing the range in predicted spring force. Considering different $N_R$ at S = 0.015, data again appear to exhibit $1/N_R^{1/2}$ scaling.

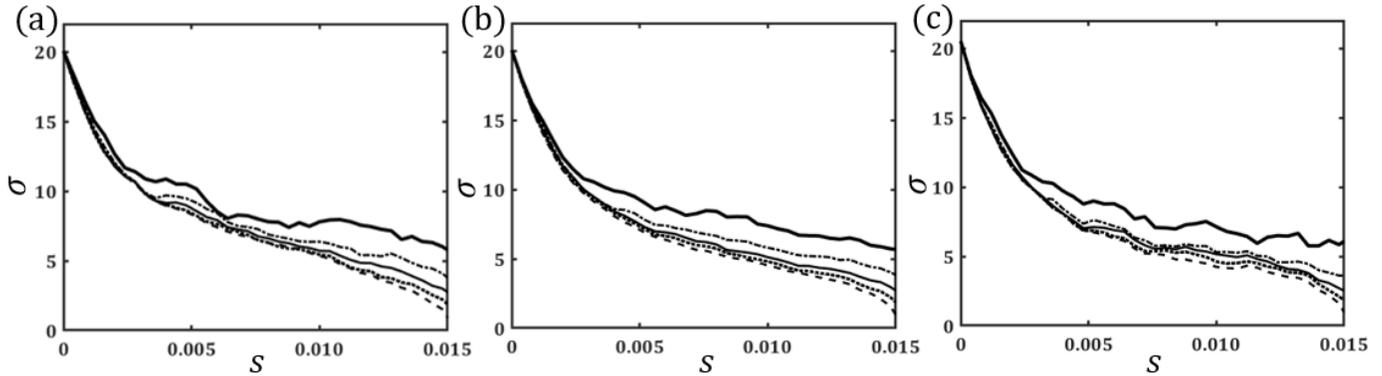

**Figure 5**: The standard deviation (σ) vs. S for the middle spring ($25^{th}$ spring) RSF for $N_R = 10, 20, 30, 40$ and 50 (top to bottom, respectively). The σ vs. S curves are estimated for conformations whose 25th spring's RSF for $N_R = 50$ at S = 0.015 is in the interval (a) $-7.5 < f_{25} < -6.5$, (b) $35.5 < f_{25} < 36.5$ and (c) $79.5 < f_{25} < 80.5$.

For polymers under transient conditions, mitigating BD algorithm-induced spurious fluctuations in spring force is essential but insufficient to predict force-induced polymer chain events accurately. It is also crucial that the sampling method uses polymer conformation samples that have changed minimally compared to the parent chain conformation. This is precisely the challenge to time-average spring force (TSF) sampling that motivates the use of replica-average spring force (RSF) sampling: the time required for a TSF prediction to become sufficiently bounded around the true average may be such that the resulting answer reflects a range in polymer conformations, some of which may differ non-negligibly from the parent chain conformation. RSF sampling allows replicas to evolve over time before an average is taken over the instantaneous spring force from each replica; thus, it is necessary to evaluate how much the parent chain conformation changes over a typical cycle of replica evolution.

We employed the results of the collapsed polymer under a high shear rate ($\dot{\gamma} = 30$) to estimate the change in replica conformations compared to their parent chain conformation. The data presented above concluded that RSF sampling should utilize a minimum evolution time on the order of S = 0.010 to S = 0.015. Thus, we evaluate the change in conformation between replica chains and their parent chain for S = 0.0075 and S = 0.0150. We computed the difference in conformation between each replica chain and its parent chain after evolution time S as:

$$\overline{D}(j, S, k) = \frac{\sum_{i=1}^{N} abs[\hat{r}_j^i(S,k) - \hat{r}_j^i(P_c)]}{N} \quad j = x, y, z \quad (5)$$

where k is the $k^{th}$ replica, $\hat{r}_j^i(S, k)$ is the $j^{th}$ component of the $i^{th}$ bead position in the polymer center of the mass frame; that is, $\hat{r}_j^i(S, k) = r_j^i(S, k) - r_j^{cm}(S, k)$, where $r_j^{cm}(S, k)$ is the $j^{th}$ component of the polymer chain center of mass at evolution time S and for replica k. Thus, $\hat{r}_j^i(P_c)$ refers to the same bead i and its same coordinate component j, but in the parent chain and also in the polymer center of mass frame. The summation over N = 50 is the number of beads in each bead-spring polymer chain.

For collapsed polymer chains under high shear rate flows, the rate of change of polymer conformation is most remarkable for chains with the most extensive variation in the shear direction z [37]. For example, if one computes the z component contributions to the chain radius of gyration, $R_{GZ}$, those chains with the largest $R_{GZ}$ exhibit the highest rates of conformational change. This often manifests in chains transitioning between the globule and elongated states, or vice-versa[35]. This suggests that $\overline{D}$ significantly depends on the z-component contributions to the parent chain conformation's radius of gyration. Here, we illustrate the distribution of $\overline{D}$ among 50 replicas (Fig. 6a) for a parent chain having the maximum value of $R_{GZ}$ exhibited by collapsed polymers at the shear rate $\dot{\gamma} = 30$. Data in Fig. 6a, therefore, illustrate the worst-case scenario as far as how much replicas may change from a parent conformation over an evolution time S. This is further illustrated in Fig. 6b, where we have shown $\overline{D}$ distribution data but for a parent chain whose $R_{GZ}$ is equal to the mean $R_{GZ}$ computed at $\dot{\gamma} = 30$.

Data in Fig. 6 for x-component contributions to $\bar{D}$ are plotted along a greater span of values than the same data for y-component and z-component contributions because the largest values in $\bar{D}$ are consistently observed for the flow direction components. It is also apparent that $\bar{D}$ values for replicas of the parent chain with the largest exhibited $R_{GZ}$ are greater than values for replicas of the parent chain with a mean value of $R_{GZ}$, but this is most obvious for the x-direction components. Computed distances are in terms of bead radius and the largest values observed ($\bar{D}(x) \sim 0.65$) are small enough to argue that the sampled replica conformations suitably represent the parent chain conformation, even after $S = 0.0150$. Figure 6c further confirms this. The top image of Fig. 6c shows the replica that exhibited the maximum value in $\bar{D}(x)$ after evolution time $S = 0.0075$; superimposed over that image is an image of that replica's parent chain, where the two chains' center of masses have been made coincident. The bottom image shows the same presentation for the replica that exhibited the maximum value in $\bar{D}(x)$ (and the parent chain), but for evolution time $S = 0.0150$. For $S = 0.0150$, and especially for $S = 0.0075$, Fig. 6c shows that the replica faithfully represents the polymer conformation exhibited by its parent chain. Given that the illustrated replicas exhibit the largest values of $\bar{D}$, it is expected that other replicas will also faithfully represent the parent chain. Examination of other replicas confirmed this.

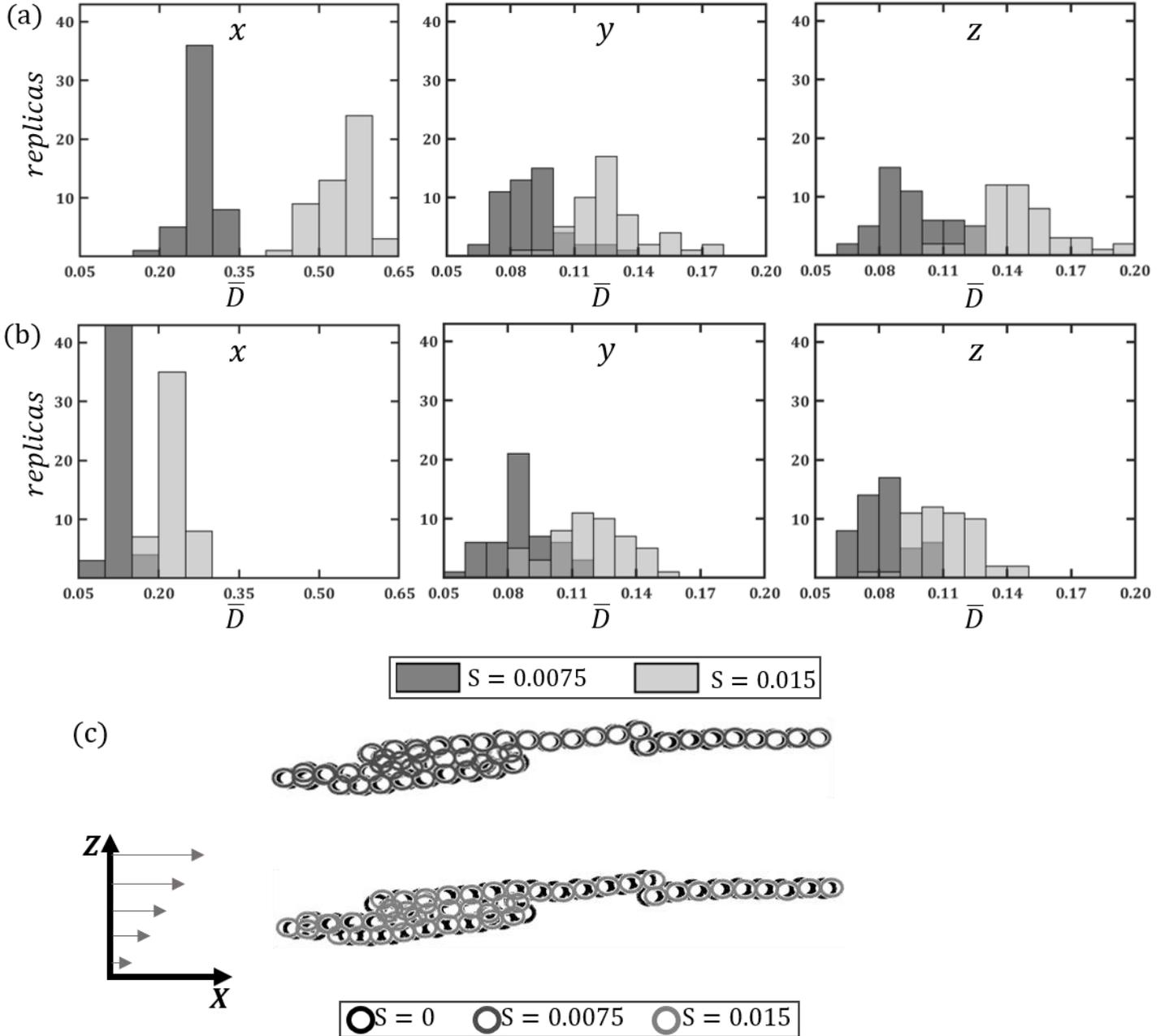

**Figure 6:** Replica chain conformation compared to parent chain conformation for polymers under high shear rate flow ($\dot{\gamma} = 30$). (a) Distribution of $\bar{D}$ over 50 replicas for a parent chain conformation having the largest observed value of $R_{GZ}$ (see text). (b) $\bar{D}$ distribution for 50 replicas, whose parent chain conformation's $R_{GZ}$ is equal to the mean value for all chains. (c) 2-D snapshot from our simulations of replica having $\bar{D}$ (j=x) maximum among $\bar{D}$ shown in (a) at S = 0.0075 (top image) and S = 0.0150 (bottom image). The parent chain snapshot is also shown with its center of mass coincident with the illustrated replica. Snapshots are collapsed into the x-z plane for clarity. The inset schematic plot in (c) shows the velocity gradient of the flow.

Careful examination of Fig. 6c, particularly the bottom image for S = 0.0150, suggests that the difference in conformation between the replica chain and parent chain is not uniform along the chain. Because data in Fig. 6 were for D normalized over all beads in a replica, i.e., $\bar{D}$, it is, therefore, essential to examine the distribution of D observed across beads and replicas. In other words, we again compute Eq. 5 but without taking an average over beads (i.e., D(j, S, k, i), where i is the $i^{th}$ bead); this analysis is only presented for the parent chain that exhibited the largest value of $R_{GZ}$. Results are shown in Fig 7, where we have only shown results for D(j = x) because for all beads, replicas, and S, D(x) is significantly greater than D(y) and D(z).

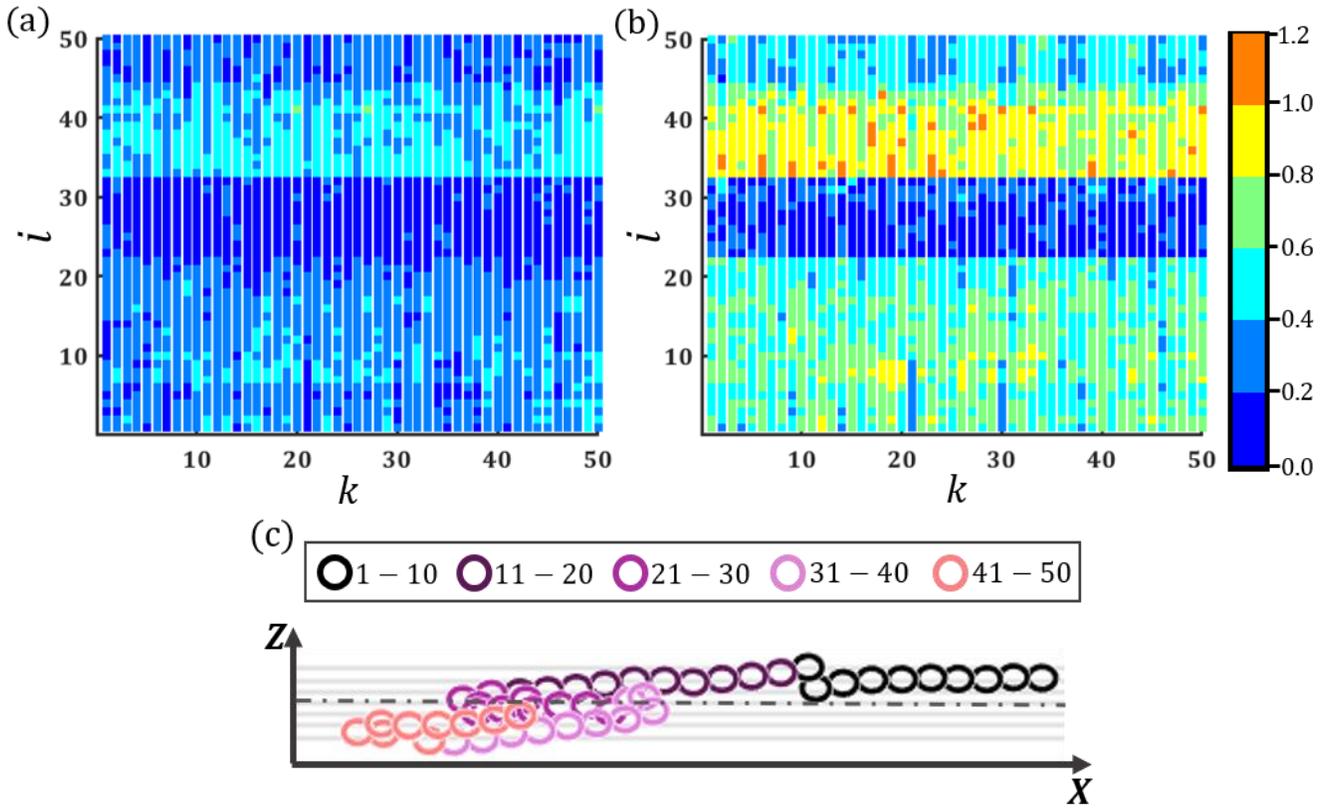

**Figure 7:** (a,b) Color map of D(j=x, k, i), where k is $k^{th}$ replica number, and i is the bead number at (a) S = 0.0075 and (b) S = 0.0150 for the parent chain conformation having the largest observed value of $R_{GZ}$. (c) 2-D snapshot of the parent chain whose replicas' D(j=x, k, i) data are shown in (a) and (b). A black dash-dotted line represents a line passing through the polymer center of mass in Z. Light gray lines are drawn such that the distance between two consecutive lines is unity.

Figure 7 illustrates D(j = x, k, i) for the replicas of the parent chain having the largest observed value of $R_{GZ}$ at S = 0.0075 and S = 0.0150. In all replicas of this parent chain, beads 33 to 41 consistently exhibit the largest values of D(j = x, k). This non-uniformity in D(j = x, i) among beads for all replicas is because, for polymer chains under shear flow, the hydrodynamic force on the polymer segments increases with the distance of segments from the polymer center of mass in the shear gradient direction, z. Figure 7(c) shows an image of the parent chain, where beads are color-coded to indicate their bead position along the chain (i.e., 1 to 50). Figure 7(c) depicts that those beads consistently showing the largest D(x) values are also farthest from the chain center of mass in z. Nevertheless, even for those beads at S = 0.0150, most have D(x) less than the bead radius.

Results demonstrate that RSF sampling over an evolution time comparable to the instantaneous force decorrelation time computed at zero flow mitigates the influence of BD algorithm-induced fluctuations in spring force and preserves the parent chain conformation over all replicas. Therefore, in highly dynamic conditions that are likely germane to mechano-reactive molecular systems, RSF sampling can predict a chain conformation's quasi-equilibrium spring force distribution more accurately than ISF and TSF sampling. The implication is that RSF sampling should more accurately predict mechano-reactive kinetics in BD simulations.

To illustrate this, a specific system is modeled: the mammalian glycoprotein von Willebrand factor (vWF), which is critical to initiating blood clotting[38]. vWF is a macromolecular polymer that is collapsed in normal blood flow conditions; vascular injury increases flow in the vicinity of the cut, which drives vWF to elongate[38,39]. This conformational change induces vWF reactivity, making it bind to platelets in blood and collagen exposed on injured vessel walls, initiating the blood clotting cascade[38–40]. vWF functionality is associated with the A2 domain in each monomer; each A2 domain is capable of unfolding, thereby exposing reaction sites[14,41]. A2 domain unfolding is force-mediated, and single-molecule force spectroscopy experiments have been previously used to advance the following rate law for A2 domain unfolding: $k = k_0 e^{f/f_u}$ where $k_0 = 7 \times 10^{-4}$s is the zero-force unfolding rate and $f_u$ is a force normalization parameter obtained from experiments $f_u = 1.1$ pN; f is the force acting on the A2 domain[41]. During a BD simulation, this rate equation can be used to predict A2 domain unfolding; however, the input force f must be free from algorithm-induced spurious fluctuations.

| Force Sampled | MFTU (seconds) |
|---|---|
| ISF | $1.82 \times 10^{-5}$ |
| TSF, S = 0.0075 (0.0150) | 0.04 (0.42) |
| RSF, $N_R = 10$, S = 0.0075 (0.0150) | 5.12 (5.98) |
| RSF, $N_R = 20$, S = 0.0075 (0.0150) | 11.69 (13.04) |
| RSF, $N_R = 40$, S = 0.0075 (0.0150) | 13.22 (13.22) |

**Table 1**: Prediction of the mean first passage time to unfold anyone A2 domain (MFTU) for vWF protein having 25 A2-domains with ISF, TSF, and RSF sampling method. For TSF S = 0.0075, 0.0150; RSF sampling method parameters are $N_R = 10, 20, 40$ and S = 0.0075, 0.0150.

Using parameters in the bead-spring model here for vWF (see Supplementary Material) for a chain with 50 beads (for our coarse grain model of vWF, this corresponds to 25 monomers or 25 A2 domains[14]), we computed the mean first passage time to observe an A2 domain unfolding (MFTU) along a chain at zero flow conditions. Based on the rate law above, the assumed vWF chain length, and considering that the average force dictates the A2 domain unfolding, the expected MFTU should be ~50 seconds[17]. With ISF sampling, Table 1 shows this time is reduced by over six orders of magnitude, invalidating any predictions of mechano-reactive kinetics. TSF sampling is appropriate in the zero flow condition, and Table 1 shows that the accuracy in the TSF predicted mean first passage time is significantly improved compared to ISF. Nonetheless, the best TSF prediction is two orders of magnitude less than the rate law predicts because of the relatively short sampling times. RSF sampling improves the situation further, with the best estimate from RSF being approximately one-fifth of the value predicted by the rate law. The zero flow case was presented here because the average force profile is well defined; nonetheless, the total time S (sampling time for TSF or evolution time for RSF) was limited similarly to how was done for high shear rate flow above. While it would be possible for the zero-flow condition to increase S for TSF to achieve similar accuracy in predicting MFTU as is achieved for RSF with $N_R = 40$ and S = 0.0150, the required sampling time would be about S = 0.4. In a high shear rate of flow, such a sampling time leads to significant changes in the parent chain conformation; for example, the x-component of $\overline{D}$ obtained at S = 0.4 are in the range 15≤ $\overline{D}(x)$ ≤18. This indicates that sampling for such times convolutes what should likely be considered structurally distinct polymer conformations.

## IV. Conclusion

A polymer chain conformation's quasi-equilibrium (average) force profile dictates the mechano-driven reaction kinetics for polymers under flow conditions[17]. We have shown that ISF fluctuations around the average force profile in BD simulations are significantly high, regardless of flow conditions. Such high fluctuation results in inaccurate estimation of mechano-reactive events kinetics if ISFs are employed to predict such events. For instance, we found that the mean first passage time for a vWF protein under quiescent flow conditions to exhibit an A2 domain unfolding computed using ISF is approximately six orders of magnitude smaller than the expected value. Therefore, sampling methods such as time-average and replica-average are required to study the influence of flow on mechano-driven polymer functionality. Results presented here show that RSF sampling over evolution time comparable to the ISF decorrelation time ensures good accuracy in predicting mechano-reactive kinetics. To achieve similar accuracy in TSF-based predictions, sampling time S, approximately ten times the ISF decorrelation time, is required. Therefore, RSF sampling should more accurately estimate the mechano-reactive events kinetics for transient conformation conditions than ISF or TSF sampling.

For the RSF sampling method, there is an optimal time beyond which simulating replicas minimally changes the standard deviation of RSF around the average force. Moreover, that optimal time is comparable to the ISF decorrelation time computed at zero flow. Further, we illustrated that when replicas are simulated for a time close to this optimal time, the replicas' chain conformation negligibly changes compared to the parent chain conformations. Therefore, the RSF sampling method accurately predicts a chain conformation's quasi-equilibrium force distribution, and this is true even for highly dynamic conditions. The standard deviation of RSF around the average value scales as $1/N_R^{1/2}$. This indicates that increasing $N_R$ beyond a particular value is not advisable considering the increment of the computational cost with $N_R$. For the collapsed polymer system studied here, the optimal number of replicas is in the range [20-40].

The present study elaborates on implementing the replica-average sampling method to study flow-induced mechano-reactive polymer functionality. Nonetheless, we believe that the replica-average sampling technique would help predict mechano-driven events in varied other processes, such as protein-collagen binding, studied using BD simulations.

**Supplementary Material**

In the Supplementary Material, Figure S1 shows the schematic diagram of the RSF sampling methodology. For polymer under no-flow conditions, Figure S2 shows the TSF's and RSF's standard deviation vs S plot for varied spring locations. Instantaneous spring force (ISF) distribution for conformation-specific spring force during high shear rate flow is illustrated in Figure S3 for the 1st and 12th springs. The parameters of the coarse-grained vWF polymer model are shown in Table S1.

**Author's Contributions**

All authors contributed to the preparation of the manuscript. S.K. performed BD simulations and data curation. A.N. assisted in manuscript revision. A.O., E.W. critically revised the manuscript and offered computational expertise.

**Conflict of Interest**

The authors have no competing interests.

**Acknowledgments**

The computational results presented have been achieved using Lehigh University's Research Computing infrastructure, partially supported by the National Science Foundation award 2019035.
# References

[1] S.S.M. Konda, S.M. Avdoshenko, and D.E. Makarov, J. Chem. Phys. **140**, 104114 (2014).

[2] D.E. Makarov, J. Chem. Phys. **144**, 30901 (2016).

[3] E. Evans and K. Ritchie, Biophys. J. **72**, 1541 (1997).

[4] G. I. Bell, Science. **200**, 618 (1978).

[5] Y. Wang, C. Zhang, E. Zhou, C. Sun, J. Hinkley, T.S. Gates, and J. Su, Comput. Mater. Sci. **36**, 292 (2006).

[6] D.E. Makarov, Acc. Chem. Res. **42**, 281 (2009).

[7] U.F. Röhrig and I. Frank, J. Chem. Phys. **115**, 8670 (2001).

[8] M.K. Beyer, J. Chem. Phys. **112**, 7307 (2000).

[9] H.J. Choi, C.A. Kim, J.I. Sohn, and M.S. Jhon, Polym. Degrad. Stab. **69**, 341 (2000).

[10] A.S. Pereira and E.J. Soares, J. Nonnewton. Fluid Mech. **179–180**, 9 (2012).

[11] J.D. Clay and K.W. Koelling, Polym. Eng. Sci. **37**, 789 (1997).

[12] B. Huisman, M. Hoore, G. Gompper, and D.A. Fedosov, Med. Eng. Phys. **48**, 14 (2017).

[13] S. Lippok, M. Radtke, T. Obser, L. Kleemeier, R. Schneppenheim, U. Budde, R.R. Netz, and J.O. Rädler, Biophys. J. **110**, 545 (2016).

[14] C. Dong, S. Kania, M. Morabito, X.F. Zhang, W. Im, A. Oztekin, X. Cheng, and E.B. Webb, J. Chem. Phys. **151**, 124905 (2019).

[15] K.D. Knudsen, J.G. Hernandez Cifre, J.J. Lopez Cascales, and J. Garcia de la Torre, Macromolecules **28**, 4660 (1995).

[16] S. Wu, C. Li, Q. Zheng, and L. Xu, Soft Matter **14**, 8780 (2018).
25

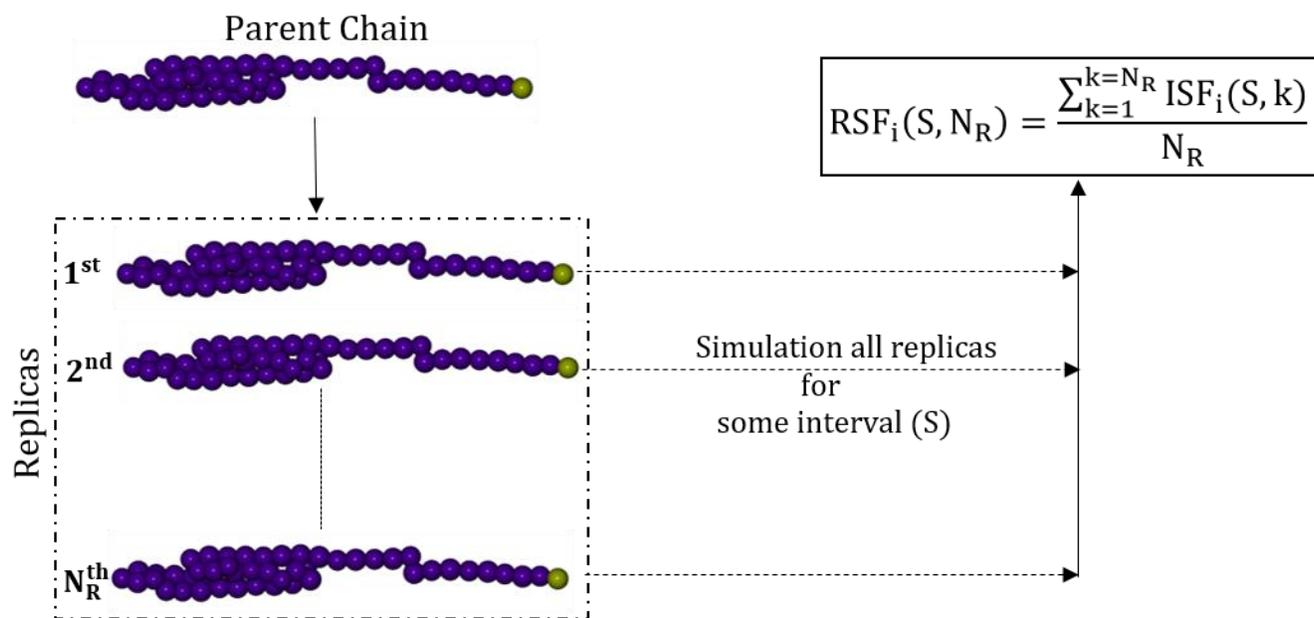

**Figure S1**: Schematic of replica-averaging sampling method procedure to compute $i^{th}$ spring RSF for evolution time S and number of replicas $N_R$; k is replica number.

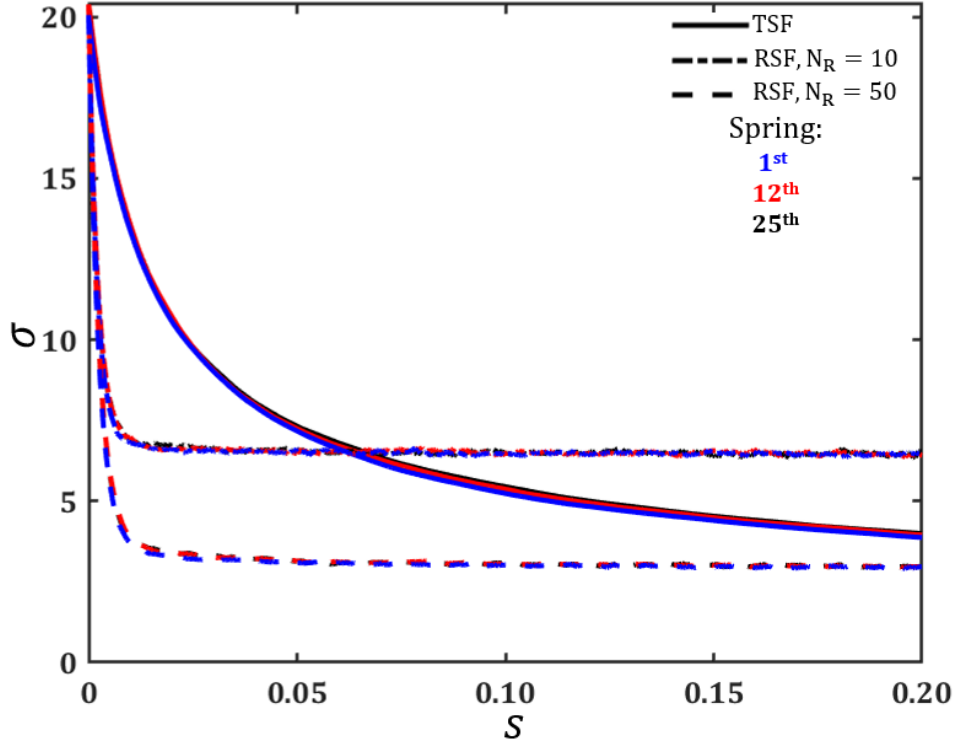

**Figure S2**: The standard deviation ($\sigma$) vs. S for RSF and TSF sampling for the $1^{st}$, $12^{th}$, $25^{th}$ spring. For RSF, data for ($N_R = 10, 50$) are shown. Data for $25^{th}$ spring is the same as in Fig. 2

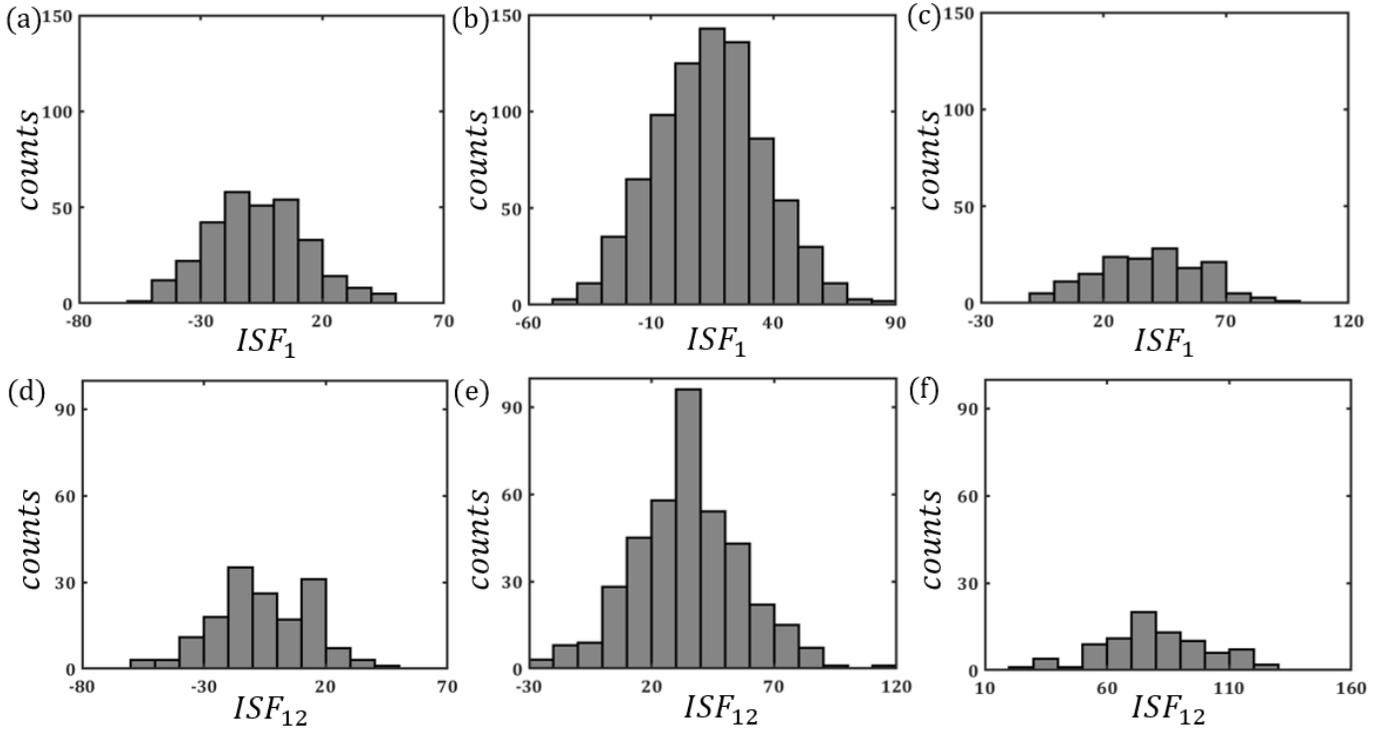

**Figure S3**: $1^{st}$ spring instantaneous force ($ISF_1$) distribution for conformations whose $1^{st}$ spring RSF for S = 0.015 and $N_R = 50$ is in the range (a) $-7.5 < f_1 < -6.5$, (b) $12.5 < f_1 < 13.5$ and (c) $35.5 < f_1 < 40.5$. Similarly,

ISF$_{12}$ distribution for conformations having 12$^{th}$ spring RSF for S = 0.015 and N$_R$ = 50 in the range (d) -7.5 < f$_{25}$ < -6.5, (e) 33.5 < f$_{25}$ < 34.5 and (f) 79.5 < f$_{25}$ < 80.5.

**Coarse-grained vWF polymer model:**

vWF monomers are modeled via a bead-spring description; this model was presented previously but is briefly summarized[1]. Two beads, connected by a parallel system of two springs, represent a monomer; the spring system consists of a finitely extensible non-linear elastic (FENE) spring and a relatively stiffer harmonic spring. The summation of those spring forces represents the force acting on the folded A2 domain. This model has been previously shown to accurately represent the complex mechanical response to pulling exhibited by A2 domains in experiment[1]. ISF, TSF, and RSF sample FENE and harmonic spring force summation to predict the A2 domain unfolding. Nonetheless, harmonic spring force dominates this summation. A bead represents all domains on either side of the A2 domain. The disulfide bond between monomers is modeled as a stiff harmonic spring. The parameter values of this model are shown in Table S1.

**Table S1.** vWF polymer model parameters with the explanation of meaning and values.

| Parameter | Meaning | Value |
|---|---|---|
| a | Bead radius | 15 [nm] |
| η | Flow viscosity | 0.001 [pN μs/nm$^2$] |
| ζ | Drag coefficient | 6πηa [pN μs/nm] |
| T | Temperature | 300 [K] |
| ε | Lennard-Jones energy parameter | 4.143 [pNnm] |
| f$_{harm}$ | Inter-monomer harmonic spring constant | 7.5 [pN/nm] |
| f$_A$ | Spring constant of FENE spring in A2 | 0.116 [pN/nm] |
| f$_B$ | Spring constant of Harmonic spring in A2 | 5 [pN/nm] |
| r$_{max}$ | Maximum length for FENE spring | 51.46 [nm] |
| k$_0$ | Zero force unfolding rate | 0.0007 [s$^{-1}$] |
| $f_u$ | Force scale | 1.1 [pN] |
| $f_{attempt}$ | State change attempt frequency | $2 \times 10^7$ [Hz] |

For vWF polymer in no-flow conditions, our simulation results indicate that the average force exerted on the A2 domain is approximately 0.28pN. Further, considering that this average force dictates the A2 domain unfolding, we computed the expected time to observe anyone A2 domain to unfold for vWF polymer consisting of 25-monomers as: $1/[25 * k_0 e^{0.28/f_u}]$ ($k_0$ is the zero-force unfolding rate and $f_u$ is a force normalization parameter), which equals 49.8 seconds. In the main manuscript, this value is approximated to 50 seconds.